\documentstyle[seceq,epsf,wrapfig,twoside]{ptptex}
\notypesetlogo  
\title{Non-leptonic Two-meson Decays of {\mib B}-Mesons
in the Covariant Oscillator Quark Model with Factorization
Ansatz}
\author{%
Rukmani {\sc Mohanta}, Anjan K. {\sc Giri}, Mohinder P. {\sc Khanna}\\
Muneyuki {\sc Ishida},$^{*}$ Shin {\sc Ishida}$^{**}$
and Masuho {\sc Oda}$^{***}$
}
\inst{%
Department of Physics,
Panjab University, Chandigarh - 160014, India\\
$^{*}$Department of Physics, Tokyo Institute of Technology\\
 Tokyo 152-8551, Japan\\
$^{**}$Atomic Energy Research Institute, 
College of Science and Technology\\
Nihon University, Tokyo 101-0062, Japan\\ 
$^{***}$Faculty of Engineering, Kokushikan University, 
Tokyo 154-8515, Japan\\
}
\recdate{%
November 20, 1998
}
\abst{%
Exclusive nonleptonic bottom meson decays are studied in
the covariant osillator quark model using the factorization
assumption. The main feature of this model is that it
can simultaniously be applied to both heavy $\to $ heavy
and heavy $\to $ light transitions, satisfying the
constraints of the heavy quark effective theory (HQET) 
in the appropriate limit.
The results obtained are in overall agreement with
the present experimental data for various $B$ decays.
}

\begin{document}
\maketitle

\setcounter{tocdepth}{4}

\section{Introduction}
The description of exclusive nonleptonic decays of $B$
mesons represents an important and complicated theoretical
problem. These decays are nonperturbative in nature
and cannot be calculated reliably from the QCD Lagrangian.
In contrast to exclusive semileptonic decays, where the weak
current matrix elements between meson states are involved,
nonleptonic decays require the evaluation of hadronic matrix
elements of four local quark operators. To simplify the analysis
it is usually assumed that  the matrix element of the 
current-current weak interaction factorizes into the product of two
single current matrix elements. Thus the problem reduces to
the calculation of the meson form factors, which are contained in the
hadronic matrix elements of weak currents as in the case of
semileptonic decays, and of the meson decay constants, describing
the leptonic decays. This makes the factorization hypothesis
\cite{ref1} a very appealing assumption. Although it is
very difficult to prove the factorization hypothesis
theoretically within our present understanding of QCD
nonperturbative effects, this hypothesis is expected to be valid to 
a rather good approximation in the case of transitions with
large energy release, such as heavy $B$ decays, since the final
mesons carrying large momenta escape from the
region of interaction, thereby minimizing the effects
of a final state interaction.
Several tests have been made to prove
its validity phenomenologically,\cite{ref2} and, it has been shown
to work well for the description of $B$
meson decays  into a $D$ or $D^*$ and a light meson.
Once the factorization assumption
is made, nonleptonic decays are related to the corresponding
semileptonic decays.

In this paper we wish to calculate the branching ratios of the
exclusive nonleptonic two-meson decays of $\bar B^0 $, $B^-$ and 
$\bar B_s$ mesons in
the framework of the covariant oscillator quark model (COQM)
\cite{ref3} on the basis of factorization.
One of the most important motivations of the covariant oscillator
quark model (COQM) is to describe covariantly the centre of
mass motion of hadrons, retaining the considerable sucesses of
the non-relativistic quark model with regard to the static properties of
hadrons. A key element of the COQM for achieving this is 
its direct treatment of 
the squared masses of hadrons, in contrast to the mass itself
as done in conventional approaches. This makes the covariant treatment
simple. The COQM has been applied to various problems
\cite{ref4} with satisfactory results. Recently, Ishida et al
have studied the semi-leptonic weak decays of heavy hadrons 
using this model
\cite{ref5} and derived the same relations of weak form
factors as in HQET.\cite{ref6} 
Furthermore, the predicted spectra for $B\rightarrow (D^*,D)l\nu$
were shown to fit experimental data quite well.\cite{spe}
Keeping this
success in mind, we extend application of the COQM to the nonleptonic
decays of $\bar B^0 $, $B^-$ and $\bar B_s $ mesons.

This paper is organized as follows. In \S 2 we present
the expressions for nonleptonic decay amplitudes in the factorization
approximation. In \S 3 we present a brief description
of the covariant oscillator quark model. Using this model
we have evaluated the form factors and obtained the decay rates
for various nonleptonic decay processes.
The decay modes $B \to D^*V $ are
considered in \S 4. Section 5 contains our results
and discussion.

\section{General formalism}
Neglecting the penguin contribution, the effective
Hamiltonian describing the decays under consideration
is given by

\begin{eqnarray}
&&{\cal H}_{\rm eff} = \frac{G_F}{\sqrt{2}}\; V_{cb}\;V_{q_i q_j}\;
[C_1(m_b) O_1 + C_2(m_b) O_2 ]
\end{eqnarray}
with
\begin{eqnarray}
&&O_1=(\bar q_i q_j)^{\mu}\; (\bar c b)_{\mu}~~~~~~~~~~
\mbox{and}~~~~~~~~~~~ O_2=
(\bar q_i b)^{\mu} (\bar c q_j)_{\mu}\;,
\end{eqnarray}
where $C_1$ and $C_2$ are the Wilson coefficients,
and the quark current $(\bar q_i q_j )_{\mu} $ denotes
the usual $(V-A)$ current. $q_i$ and $q_j $  are two types
of quark flavors that are hadronized to the $P$ or $V$
mesons.

The factorization approach to two-body nonleptonic decays
$B \to D M $ implies that the decay amplitude can be
expressed by the product of one particle matrix elements:
\begin{eqnarray}
\langle D M | {\cal H}_{\rm eff} | B \rangle &=& \frac{G_F}{\sqrt 2}
V_{cb} V_{q_i q_j}~\biggr[ a_1 \langle D |
(\bar c b)_{\mu} |B \rangle \langle
M |(\bar q_i q_j )^{\mu} | 0 \rangle \nonumber\\
& & +  a_2 \langle M |(\bar q_i b)_{\mu} |B \rangle \langle
D |(\bar c q_j )^{\mu} | 0 \rangle \biggr] \ .
\label{eq:3}
\end{eqnarray}
Here $a_1 = C_1 + C_2/N_c $ and $a_2=C_2 + C_1/N_c$, where $N_c$
represents the number of colors.

It should be noted that in writing Eq.~(\ref{eq:3}) we have discarded the
contribution of color octet currents which emerges after the Fierz
rearrangement of color singlet operators. Clearly these currents
violate factorization since they cannot allow transition to
the vacuum states. In the factorization approximation one can
distinguish\footnote{
The contributions due to quark annihilation processes,
which are expected to be small,\cite{ref1} are neglected in the 
calculation of this paper.
} 
three classes of $B$ meson decays : the `class I'
transitions such as $\bar B^0 \to M_1^+ M_2^- $, where only
the term $a_1$ contributes (both mesons produced by charged
currents ); `class II' transitions, such as $ \bar B^0 \to
M_1^0 M_2^0 $, where only the term $a_2$ contributes (both mesons
are produced by neutral currents); and `class III' transitions,
such as $B^- \to M_1^0 M_2^- $, where both terms contribute
coherently.

In order to evaluate the transition amplitudes we use the following
matrix elements:
\begin{eqnarray}
&&\langle P(p) |(\bar q_i q_j)^{\mu} |0 \rangle
= -if_P~ p^\mu , \nonumber\\
&&\langle V(p,\epsilon) |(\bar q_i q_j)^{\mu} |0 \rangle
= M_V~f_V~ \epsilon^\mu , \nonumber\\
&&\langle a_1(p,\epsilon) |(\bar q_i q_j)^{\mu} |0 \rangle
= M_{a_1}~f_{a_1}~ \epsilon^\mu\;,
\label{eq:4}
\end{eqnarray}
where $P$, $V$ and $a_1 $ represent the pseudoscalar, vector
and the axial vector mesons, respectively.
To evaluate the hadronic form factors we use the COQM.
These are presented in the next section.

\section{ Model Framework, hadronic form factors\\
and decay width of B$\to$ PP, B$\to$ PV and B$\to$ VP}

The general treatment of COQM may be called
the ``boosted $LS$-coupling scheme,'' and the wavefunctions being
tensors in $\tilde U(4) \times O(3,1) $-space, reduce to
those in $SU(2)_{\rm spin} \times O(3)_{\rm orbit} $-space in the
nonrelativistic quark model in the hadron rest frame. The
spinor and space-time portion of the wave functions
separately satisfy the respective covariant equations,
the Bargmann-Wigner (BW) equation for the former and the
covariant oscillator equation for the latter. The form of
the meson wave function has been determined completely
through the analysis of mass spectra.

In COQM, the meson states are described by bi-local fields
$\Phi_A^B(x_{1\mu},x_{2\mu}) $, where $x_{1\mu}(x_{2\mu})$ is
the space time coordinate of the constituent quark (antiquark),
$A=(a,\alpha)\;(B=(b,\beta))$ describing its flavor and
covariant spinor. Here we write only the positive frequency
part of the relevant ground state fields:
\begin{equation}
\Phi_A^B(x_{1\mu},x_{2\mu})=e^{iP\cdot X}\; U(P)_A^B\;
f_{ab}(x_{\mu};P)\;,
\end{equation}
where $U$ and $f$ are the covariant spinor and internal
space-time wave functions respectively, satisfying the
Bargmann-Wigner and oscillator wave equations. The quantity
$x_{\mu} (X_{\mu}) $ is the relative (CM) coordinate,
$x_{\mu}\equiv x_{1\mu}-x_{2\mu} (X_{\mu}\equiv m_1x_{1\mu}
+m_2x_{2\mu})/(m_1+m_2) $, where the $m_i$ represent the quark
masses). The function $U$ is given by
\begin{equation}
U(P)=\frac{1}{2 \sqrt 2}\left [(-\gamma_5 P_s(v)+i\gamma_{\mu}
V_{\mu}(v))(1+iv \cdot \gamma)\right ] ,
\end{equation}
where $P_s(V_s)$ represents the pseudoscalar (vector) meson
field, and $v_{\mu} \equiv P_{\mu}/M$ [$P_{\mu}(M)$ is the
four momentum (mass) of the meson]. The function $U$, being
represented by the direct product of quark and antiquark
Dirac spinors with the meson velocity, is reduced to the
non-relativistic Pauli-spin function in the meson rest frame.
The function $f$ is given\footnote{
In this paper we employ the pure-confining approximation,
neglecting the effect of the one-gluon-exchange potential,
which is expected to be good for the heavy/light-quark meson system.
} by
\begin{equation}
f(x_{\mu};P)=\frac{\beta}{\pi} \exp\left (-\frac{\beta}{2}
\left (x_{\mu}^2+2\frac{(x\cdot P)^2}{M^2}\right ) \right )\; .
\end{equation}
The value of the parameter $\beta $ is determined from the mass
spectra \cite{ref8} as $\beta_{\pi/\rho/a_1}=0.14 $, $\beta_{K/K^*}$
=0.142, $\beta_{D/D^*} $=0.148, $\beta_{D_s}=0.154 $, $\beta_B $
=0.151 and $\beta_{B_s} $=0.160 (in units of
$ \mbox{GeV}^2$ ).

The effective action for weak interactions of mesons
with $W$-bosons is given by
\begin{equation}
S_W=\int d^4 x_1 d^4 x_2 \langle \bar \Phi_{F, P^\prime}
(x_1,x_2) i\gamma_{\mu}(1+\gamma_5) \Phi_{I, P}(x_1,x_2)
\rangle W_{\mu,q}(x_1)\;,\label{eq:eqn1}
\end{equation}
where we have denoted the interacting (spectator) quarks as
1 (2). The CKM matrix elements and the coupling constant
are omitted. This equation is obtained from consideration of
Lorentz covariance, assuming a quark current with the
standard $V-A$ form. In Eq. (\ref{eq:eqn1}), $\Phi_{I,P}~
(\bar \Phi_{F, P^\prime}) $ denotes the initial (final)
meson with definite four momentum $P_{\mu} (P_{\mu}^{\prime})$,
and $q_{\mu}$ is the momentum of $W$ boson. The function
$\bar \Phi $ is defined by $\bar \Phi = -\gamma_4 \Phi^\dagger
\gamma_4$, and $\langle~~\rangle $ represents the trace of
Dirac spinor indices. Our relevant effective current
$J_{\mu}(X)_{P^\prime,P} $ is obtained by identifying
the above equation with

\begin{equation}
S_W =\int d^4 X J_{\mu}(X)_{P^\prime,P}\;W_{\mu}(X)_q\;.
\end{equation}
Then $J_{\mu}(X=0)_{P^\prime,P} \equiv J_{\mu} $ is
explicitly given as \cite{ptp93}

\begin{eqnarray}
J_{\mu} = I^{qb}(w) \sqrt{M M^\prime}
&\times & [\bar P_s(v^\prime)P_s(v)(v+v^\prime )_{\mu}\nonumber\\
&+& \bar V_{\lambda}(v^\prime ) P_s(v) (\epsilon_{\mu \lambda
\alpha \beta} v_{\alpha}^{\prime}v_\beta-\delta_{\lambda \mu}
(w+1) - v_{\lambda}v_{\mu}^{\prime}]\;,
\label{eq:10}
\end{eqnarray}
where $M (M^\prime) $ denotes the physical masses of the
initial (final) mesons. It should be noted that in the pure
confining limit, the masses of the ground state mesons are
equal to the simple sums of their constituents, which are
much different from the physical masses in the case of light
quark pseudoscalar mesons, such as $\pi $ and $K$. Therefore
we do not consider the transitions $ B \to \pi $
and $B \to K $ in our analysis as the reliabilty of the results
is less for these transitions.
The quantity $I^{qb}(w) $, which is the overlapping of the initial
and final wave functions, represents the universal form
factor. It describes the confined effects
of quarks and is given by

\begin{equation}
I^{qb}(w)=\frac{4 \beta \beta^\prime}{\beta+\beta^\prime }
\frac{1}{\sqrt{C(w)}} \exp(-G(w))\;;~~~~~~~~~~~~
C(w)=(\beta-\beta^{\prime})^2+4\beta \beta^{\prime} w^2\;,
\end{equation}
and
\begin{eqnarray}
G(w)&=& \frac{m_n^2}{2 C(w)} \biggr[ (\beta+\beta^\prime)
\left \{ \left (\frac{M}{M_s} \right )^2 + \left (\frac{M^\prime}
{M_s^\prime}\right )^2 -2 \frac{M M^\prime}{M_s M_s^\prime}w
\right \}\nonumber\\
& & + 2 \left \{ \beta^\prime\left (\frac{M}{M_s} \right )^2
+\beta\left ( \frac{M^\prime}{M_s^\prime}\right )^2
\right \}(w^2-1) \biggr]\;,
\end{eqnarray}
where $M_s(M_s^\prime)$ represents the sum of the constituent
quark masses of the initial (final) meson, and $m_n $ is the
spectator quark mass.

The form factor function $I^{cb}(w)$ for $ B \to D(D^*) $decays
corresponds to the Isgur-Wise function $\xi(w) $ in HQET.\cite{ref6}
At the zero recoil point $w=1$,
the value of $I^{cb}(w)$ is given
by
\begin{equation}
I^{cb}(w=1)=\frac{4 \beta \beta^\prime}{(\beta+\beta^\prime)^2}\;.
\label{eq:13}
\end{equation}
In the heavy quark symmetry limit $\beta=\beta^\prime $, so
Eq.~(\ref{eq:13}) correctly reproduces\cite{spe} 
the normalization condition of
HQET, i.e., $\xi(w=1)=1 $. However, HQET, as it is, predicts
nothing about the Isgur-Wise function except for the zero recoil
point, while in COQM the form factor functions can be derived
at any kinematical point of interest due to the fact that the 
center of mass
motion of the meson there is treated covariantly,
as was mentioned in \S 1.
In addition, the COQM form factor $I^{qb}(w)$ is
also applicable for the heavy-to-light transition processes,
while HQET does not provide anything for this sector.

After obtaining the effective current in the COQM, the decay widths
for various $B\to D$ and $B\to D^*$ decay modes can be obtained with 
Eqs. (\ref{eq:3}), (\ref{eq:4}) and (\ref{eq:10}). 
These are as follows:  

\begin{eqnarray}
\Gamma(B(v) &\to &  D(v_1) P(v_2)) = 
\frac{G_F^2}{16 \pi M_{B}^2}~
~|V_{cb} V_{q_i q_j} |^2~|{\bf p}|\nonumber\\
& \times & \biggr[a_1~f_P~\sqrt{M_{B} M_D}(I_q^{cb}(w_1))
(1+w_1)~\left (M_B-M_D\right ) \nonumber\\
& + & a_2~f_D~\sqrt{M_{B} M_P}(I_q^{qb}(w_2))
(1+w_2)~\left (M_B-M_P\right )\biggr]^2\;,
\end{eqnarray}

\begin{eqnarray}
\Gamma(B(v) &\to &  D^*(v_1) P(v_2)) = 
\frac{G_F^2}{16 \pi M_{B}^2}~
~|V_{cb} V_{q_i q_j} |^2~|{\bf p}|^3\nonumber\\
&\times & \biggr[a_1~f_P~\sqrt{\frac{M_B}{ M_{D^*}}}(I_q^{cb}(w_1))
~\left (M_B+M_{D^*}\right )\nonumber\\
&+& a_2~f_{D^*}~\sqrt{\frac{M_B}{ M_P}}(I_q^{qb}(w_2))
~\left (M_B+M_P\right )\biggr]^2\;,
\end{eqnarray}

\begin{eqnarray}
\Gamma(B(v) &\to &  D(v_1) V(v_2)) = 
\frac{G_F^2}{16 \pi M_{B}^2}~
~|V_{cb} V_{q_i q_j} |^2~|{\bf p}|^3\nonumber\\
&\times & \biggr[a_1~f_V~\sqrt{\frac{M_B}{ M_D}}(I_q^{cb}(w_1))
~\left (M_B+M_D\right )\nonumber\\
&+& a_2~f_D~\sqrt{\frac{M_B}{ M_V}}(I_q^{qb}(w_2))
~\left (M_B+M_V\right )\biggr]^2\;,
\end{eqnarray}
where we have taken $w_1 = v \cdot v_1 $ and $w_2=v \cdot v_2 $.
Here $|{\bf p}| $ is the c.m. momentum of the emitted particles
in the rest frame of the $B$ meson.
As stated earlier, only the $a_1 $ term contributes to class I
decays, and only the $a_2 $ term contributes to class II decays,
while both $a_1 $ and $a_2 $ terms 
contribute coherently to class III decays.

The COQM is also applicable to heavy to light transitions,
as well as heavy to heavy transitions, such as $B\to\rho$
and $B\to K^*$. In this case the above formulas are changed by replacing 
$(D^*(v_1),I_q^{cb}(w_1),V_{cb})$ in $B\to D^*$ transition
with $(\rho (v_1),I_q^{nb}(w_1),V_{ub})$ and 
$(K^*(v_1),I_q^{sb}(w_1),V_{ub})$ in $B\to\rho$ and $B\to K^*$ 
transitions, respectively.

\section{ Decay rate, polarization and angular correlation
in the decays $B \to VV $ }

The helicity amplitude for the decay process
$ B(p) \to D^{*}(k_1,\epsilon_1)V(k_2,\epsilon_2) $
can be expressed by three invariant amplitudes, $a,~b$ and $c$. It
is given following Ref. \citen{ref9} as
\begin{equation}
H_{\lambda}=\epsilon_{1\mu}^{(\lambda )*} 
\epsilon_{2\nu}^{(\lambda )*}\left [a g^{\mu \nu}
+\frac{b}{M_{D^*}M_V} p^{\mu} p^{\nu} + \frac{ic}{M_{D^*}M_V}
\epsilon^{\mu \nu \alpha \beta}
k_{1\alpha} p_{\beta} \right ]\;.
\end{equation}
The coefficients $a$,
$b$ and $c$ describe the $S$-, $P$- and $D$- wave contributions
to the two final vector particles. In the present framework
these are given as
\begin{eqnarray}
a&=& \frac{G_F}{\sqrt 2} V_{cb} V_{q_i q_j} [
a_1 f_V M_V \sqrt{M_B M_{D^*}} ~I^{cb}(w_1)(1+w_1)\nonumber\\
 & & ~~~~~~~~~~~~+a_2 f_{D^*} M_{D^*} \sqrt{M_B M_V} ~I^{qb}(w_2) (1+w_2)
 ],\nonumber\\
b&=& -\frac{G_F}{\sqrt 2} V_{cb} V_{q_i q_j}\left [
a_1 f_V M_V^2 \sqrt{\frac{ M_{D^*}}{M_B}} ~I^{cb}(w_1)
+a_2 f_{D^*} M_{D^*}^2 \sqrt{\frac{M_V}{M_B}} ~I^{qb}(w_2)
\right ] ,\nonumber\\
c&=&- \frac{G_F}{\sqrt 2} V_{cb} V_{q_i q_j}\left [
a_1 f_V M_V^2 \sqrt{\frac{ M_{D^*}}{M_B}} ~I^{cb}(w_1)
+a_2 f_{D^*} M_{D^*}^2 \sqrt{\frac{M_V}{M_B}} ~I^{qb}(w_2)
\right ] .\ \ \ \ \ \ 
\end{eqnarray}      
The helicity amplitudes are
given as
\begin{equation}
H_{\pm 1}=a\pm \sqrt{x^2-1}~c ~~~~~~~~\mbox{and}
~~~~~~~~~H_0=-ax-b(x^2-1) \;,
\end{equation}
where $x$ is defined by
\begin{equation}
x\equiv \frac{k_1 \cdot k_2}{M_{D^*} M_V}=\frac{M_B^2 -M_{D^*}^2
-M_V^2}{2M_{D^*} M_V}
\end{equation}
and obeys
\begin{equation}
x^2 = 1+ \frac{M_B^2 |{\bf p}|^2}{M_{D^*}^2 M_V^2}\;.
\end{equation}
The corresponding decay rate can be obtained as
\begin{equation}
\Gamma(B \to D^* V)= \frac{|{\bf p}|}{8 \pi M_B^2}
\biggr[2|a|^2 +|xa+(x^2-1)b|^2+2(x^2-1)|c|^2 \biggr]\;.
\end{equation}
The decay distribution
is parametrized by the coefficients

\begin{eqnarray}
&&\frac{\Gamma_T}{\Gamma}=\frac{|H_{+1}|^2+|H_{-1}|^2}
{|H_0|^2+|H_{+1}|^2+|H_{-1}|^2},~~~~~~~~~~~~
\frac{\Gamma_L}{\Gamma}=\frac{|H_{0}|^2}
{|H_0|^2+|H_{+1}|^2+|H_{-1}|^2},\nonumber\\
&&\nonumber\\
&&\alpha_1= \frac{\mbox{Re}~(H_{+1}H_0^*+H_{-1}H_0^*)}
{|H_0|^2+|H_{+1}|^2+|H_{-1}|^2},~~~~~~~~~~~~~
\beta_1=\frac{\mbox{Im}~(H_{+1}H_0^*-H_{-1}H_0^*)}
{|H_0|^2+|H_{+1}|^2+|H_{-1}|^2},\nonumber\\
&&\nonumber\\
&&\alpha_2=\frac{\mbox{Re}(H_{+1}H_{-1}^*)}
{|H_0|^2+|H_{+1}|^2+|H_{-1}|^2},~~~~~~~~~~~
\beta_2=\frac{\mbox{Im}(H_{+1}H_{-1}^*)}
{|H_0|^2+|H_{+1}|^2+|H_{-1}|^2}.
\end{eqnarray}

In general, the dominant terms in the angular correlations are
$\Gamma_T/\Gamma $, $\Gamma_L/\Gamma $, $\alpha_1 $
and $\alpha_2 $. The terms $\beta_1 $ and $\beta_2 $ are small
since they are nonvanishing only if the helicity amplitudes
$H_{+1} $, $H_{-1} $ and $H_0$ or the invariant amplitudes
$a$, $b$ and $c$, respectively, have different phases.

In the case of heavy to light transitions, $B\to \rho$
and $B\to K^*$, the corresponding formulas are obtained by the 
procedure explained as the end of the last section.

\section{Results and conclusion}

\begin{table}

\caption{ Branching ratios of nonleptonic $\bar B^0 $
decays in the COQM.
Note that the values do not include the possible contribution
from the penguin diagram,
which is generally expected to be of order $10^{-6}\sim 10^{-7}$.
It is shown that the penguin diagram does not contribute to
$\rho^+D_s^-,\rho^+D_s^{*-},\bar K^{*0}\bar D^0$ and 
$\bar K^{*0}\bar D^{*0}$. The contribution to 
$\rho^+\pi^-,\rho^+\rho^-$ and $\rho^+a_1^-$ is about
$3\times 10^{-7}$.$^{1)}$
} 

\begin{center}

\begin{tabular}{ccc}
\hline
\multicolumn{1}{c}{$\bar B^0 $ modes}&
\multicolumn{1}{c}{This work}&
\multicolumn{1}{c}{Expt. [15]}\\
\hline
\multicolumn{1}{c}{Class I} &
\multicolumn{1}{c}{} &
\multicolumn{1}{c}{}\\
\hline
$D^+ \pi^-$& 3.00 $\times 10^{-3} $ & $(3.0\pm 0.4)
\times 10^{-3} $\\
$D^+ K^-$& $ 2.28 \times 10^{-4}$ &- \\
$D^+ D^-$& $ 4.345 \times 10^{-4} $ &-\\
$D^+ D_s^-$& $ 10.9 \times 10^{-3} $ &
$(8.0 \pm 3.0) \times 10^{-3} $\\
$D^+ \rho^-$& $7.406 \times 10^{-3} $ &
$ (7.9 \pm 1.4) \times 10^{-3} $\\
$D^+ K^{*-}$& $ 3.84 \times 10^{-4} $ &-\\
$D^+ D^{*-}$& 3.29 $\times 10^{-4} $ &
$ < 1.2 \times 10 ^{-3} $\\
$D^+ D_s^{*-}$& $0.86 \times 10^{-2} $  &
$ (1.0 \pm 0.5 ) \times 10^{-2} $\\
$D^+ a_1^-$ & $6.51 \times 10^{-3} $ &
$(6.0 \pm 3.3) \times 10^{-3} $\\
$D^{*+} \pi^- $ & $3.07 \times 10^{-3} $ &
$(2.76 \pm 0.21 ) \times 10^{-3} $\\
$D^{*+} K^-$& $ 2.28 \times 10^{-4} $
& - \\
$D^{*+} D^-$& $3.14 \times 10^{-4} $ &
$ <1.8 \times 10^{-3} $\\
$D^{*+} D_s^-$& $7.615 \times 10^{-3} $ &
$(9.6 \pm 3.4) \times 10^{-3} $\\
$D^{*+} \rho^-$& $8.91 \times 10^{-3} $ &
$(6.7 \pm 3.3 ) \times 10^{-3} $\\
$D^{*+} K^{*-}$& $ 4.89 \times 10^{-4} $&-\\
$D^{*+} D^{*-}$&$ 8.74 \times 10^{-4} $ &
$< 2.2 \times 10^{-3} $\\
$D^{*+} D_s^{*-}$& $2.49 \times 10^{-2} $ &
$(2.0 \pm 0.7) \times 10^{-2} $\\
$D^{*+}a_1^- $ & $ 0.99 \times 10^{-2} $&
$(1.30 \pm 0.27 ) \times 10^{-2} $ \\
$\rho^+ D_s^{-}$ & $2.17\times 10^{-5}$ &
$< 7 \times 10^{-4} $\\
$\rho^+ D_s^{*-}$ & $4.63\times 10^{-5}$ &
$< 8 \times 10^{-4} $\\
$\rho^+ \pi^{-}$ & $6.53\times 10^{-6}$ &
$< 8.8 \times 10^{-5} $\\
$\rho^+ \rho^{-}$ & $1.83\times 10^{-5}$ &
$< 2.2 \times 10^{-3} $\\
$\rho^+ a_1^{-}$ & $1.94\times 10^{-5}$ &
$< 3.4 \times 10^{-3} $\\
\hline
\multicolumn{1}{c}{Class II} &
\multicolumn{1}{c}{} &
\multicolumn{1}{c}{} \\
\hline
$D^0 \rho^0$& $ 0.649 \times 10^{-4} $ &
$< 3.9 \times 10^{-4} $\\
$D^{*0}\rho^0$& $ 1.22 \times 10^{-4} $ &
$ < 5.6 \times 10^{-4} $\\
$\bar K^{*0} J/\psi $ &$1.504 \times 10^{-3}$ &
$(1.35 \pm 0.18) \times 10^{-3} $ \\
$\bar K^{*0} \bar D^{0}$ &$1.16 \times 10^{-6}$ &
$- $ \\
$\bar K^{*0} \bar D^{*0}$ &$2.25 \times 10^{-6}$ &
$- $ \\
\hline
\end{tabular}
\end{center}
\end{table}

\begin{table}

\caption{ Branching ratios of nonleptonic $B^- $
decays in the COQM. The numbers for $D^0a_1^-$,
$D^{*0}a_1^-$,$D^0\pi^-$,$D^{*0}\pi^-$,$D^0K^-$,$D^{*0}K^-$,
$\rho^0\pi^-$ and $\rho^0a_1^-$ in class III do not contain the 
contributions from the color-suppressed $a_2$-term
in Eq.~(2.3).
Note also that the values do not include the possible contribution
from the penguin diagram,
which are generally expected to be of order $10^{-6}\sim 10^{-7}$.
It is shown that the penguin diagram does not contribute to 
$\rho^0D_s^-,\rho^0D_s^{*-},\rho^-J/\psi ,K^{*-}\bar D^0$ and 
$K^{*-}\bar D^{*0}$. The contribution to 
$\rho^0\pi^-,\rho^0\rho^-$ and $\rho^0a_1^-$ is about
$3\times 10^{-7}$.$^{1)}$  
}
\begin{center}
\begin{tabular}{ccc}
\hline
\multicolumn{1}{c}{$B^- $ modes}&
\multicolumn{1}{c}{This work}&
\multicolumn{1}{c}{Expt. [15]}\\
\hline
\multicolumn{1}{c}{Class I} &
\multicolumn{1}{c}{} &
\multicolumn{1}{c}{}\\
\hline
$D^0 D^-$& 4.53 $\times 10^{-3} $ & -\\
$D^0 D_s^-$& $ 1.16 \times 10^{-2}$ &
$(1.3 \pm 0.4) \times 10^{-2} $ \\
$D^0 D^{*-}$& $ 3.48 \times 10^{-4} $ &-\\
$D^0 D_s^{*-}$& $ 9.07 \times 10^{-3} $ &
$(9.0 \pm 4.0) \times 10^{-3} $\\
$D^{*0} D^-$& $3.34 \times 10^{-4} $ &-\\
$D^{*0} D_s^{-}$& 0.81 $\times 10^{-2} $ &
$ ( 1.2\pm 0.5) \times 10 ^{-2} $\\
$D^{*0} D^{*-}$& $9.23 \times 10^{-4} $  & -\\
$D^{*0} D_s^{*-}$ & $2.625 \times 10^{-2} $ &
$(2.7 \pm 1.0) \times 10^{-2} $\\

$\rho^0 D_s^{-}$ & 1.15 $\times 10^{-5} $ &
$ < 4 \times 10 ^{-4} $\\
$\rho^0 D_s^{*-}$ & $2.45 \times 10^{-5} $ &
$ < 5 \times 10^{-4} $\\

\hline
\multicolumn{1}{c}{Class II} &
\multicolumn{1}{c}{} &
\multicolumn{1}{c}{} \\
\hline
$K^{*-} J/\psi $ &$1.587 \times 10^{-3} $
& $(1.47 \pm 0.27) \times 10^{-3} $ \\
$\rho^- J/\psi $ &$5.79 \times 10^{-5} $
& $ < 7.7 \times 10^{-4} $ \\
$K^{*-} \bar D^0 $ & $1.23 \times 10^{-6} $
& $-$ \\
$K^{*-} \bar D^{*0} $ & $2.38 \times 10^{-6} $
& $-$ \\
\hline
\multicolumn{1}{c}{Class III} &
\multicolumn{1}{c}{} &
\multicolumn{1}{c}{} \\
\hline
$D^{0} \rho^-$& $ 1.004 \times 10^{-2} $
& $ (1.34 \pm 0.18) \times 10^{-2} $ \\
$D^{*0} \rho^-$& $1.264 \times 10^{-2} $ &
$(1.55 \pm 0.31) \times 10^{-2} $\\
$D^{0} a_1^-$& $6.89 \times 10^{-3} $ &
$(5.0 \pm 4.0 ) \times 10^{-3} $\\
$D^{*0} a_1^{-}$& $ 1.04 \times 10^{-2} $&
$(1.9 \pm 0.5 ) \times 10^{-2} $\\
$D^{0} \pi^-$& $3.18 \times 10^{-3} $ &
$(5.3 \pm 0.5 ) \times 10^{-3} $\\
$D^{*0} \pi^{-}$& $ 3.25 \times 10^{-3} $&
$(4.6 \pm 0.4 ) \times 10^{-3} $\\
$D^{0} K^-$& $2.41 \times 10^{-4} $ &
$-$\\
$D^{0} K^{*-}$& $5.34 \times 10^{-4} $ &
$-$\\
$D^{*0} K^-$& $2.42 \times 10^{-4} $ &
$-$\\
$D^{*0} K^{*-}$& $7.13 \times 10^{-4} $ &
$-$\\
$\rho^{0} \pi^-$& $3.45 \times 10^{-6} $ &
$ < 4.3 \times 10^{-5} $\\
$\rho^{0} a_1^-$& $1.03 \times 10^{-5} $ &
$ < 6.2 \times 10^{-4} $\\
$\rho^{0} \rho^-$& $1.45 \times 10^{-5} $ &
$ < 1.0 \times 10^{-3} $\\
\hline
\end{tabular}
\end{center}
\end{table}

\begin{table}
\caption{ Branching ratios of nonleptonic $\bar B_s^0 $
decays in the COQM model.}
\begin{center}
\begin{tabular}{ccc}
\hline
\multicolumn{1}{c}{$\bar B_s^0 $ modes}&
\multicolumn{1}{c}{This work}&
\multicolumn{1}{c}{Expt. [15]}\\
\hline
$D_s^+ \pi^-$& 0.248 $\times 10^{-2} $ & $ < 13 \% $\\
$D_s^+ K^-$& $ 1.89 \times 10^{-4}$ &- \\
$D_s^+ D^-$& $ 3.73 \times 10^{-4} $ &-\\
$D_s^+ D_s^-$& $ 9.57 \times 10^{-3} $ &-\\
$D_s^+ \rho^-$& $6.15 \times 10^{-3} $ &-\\
$D_s^+ K^{*-}$& $ 3.196 \times 10^{-4} $ &-\\
$D_s^+ D^{*-}$& 2.86 $\times 10^{-4} $ &-\\
$D_s^+ D_s^{*-}$& $7.48 \times 10^{-3} $  &-\\
$D_s^+ a_1^-$ & $4.86 \times 10^{-3} $ & -\\
$D_s^{*+} \pi^- $ & $2.55 \times 10^{-3} $ &-\\
$D_s^{*+} K^-$& $ 1.9 \times 10^{-4} $ &-\\
$D_s^{*+} D^-$& $2.735 \times 10^{-4} $ &-\\
$D_s^{*+} D_s^-$& $6.68 \times 10^{-3} $ &-\\
$D_s^{*+} \rho^-$& $7.49 \times 10^{-3} $ &-\\
$D_s^{*+} K^{*-}$& $ 4.1 \times 10^{-4} $&-\\
$D_s^{*+} D^{*-}$&$ 6.85 \times 10^{-4} $ &-\\
$D_s^{*+} D_s^{*-}$& $2.216 \times 10^{-2} $ &-\\
$D_s^{*+}a_1^- $ & $ 8.39 \times 10^{-3} $&-\\
\hline
\end{tabular}
\end{center}
\end{table}

\begin{table}
\caption{ Polarization and angular correlation parameter
of $B \to D^* V $ decays.}
\begin{center}
\begin{tabular}{cccc}
\hline
\multicolumn{1}{c}{Decay modes}&
\multicolumn{1}{c}{$\Gamma_L/\Gamma $}&
\multicolumn{1}{c}{$\alpha_1$}&
\multicolumn{1}{c}{$\alpha_2$}\\
\hline
$\bar B^0 \to D^{*+} \rho^-$& 0.883 &-0.416 & 0.04 \\
$\bar B^0 \to D^{*+} D^{*-}$& 0.538 & -0.662 & 0.176\\
$\bar B^0 \to D^{*+} D_s^{*-}$& 0.515 & -0.665 & 0.187\\
$\bar B^0 \to \bar K^{*0}J/\psi $ & 0.428 & -0.605 & 0.142\\
$B^- \to D^{*0} \rho^-$& 0.855 &-0.446 & 0.044 \\
$B^- \to D^{*0} D^{*-}$& 0.538 & -0.661 & 0.176\\
$B^- \to D^{*0} D_s^{*-}$& 0.515 & -0.665 & 0.187\\
$B^- \to K^{*-}J/\psi $ & 0.428 & -0.605 & 0.141\\
\hline
\end{tabular}
\end{center}
\end{table}

In order to make the numerical estimate, we use the
following values of various quantities. The quark masses
(in GeV) are taken as $m_u=m_d=0.4$, $m_s=0.55$, $m_c=1.7$
and $m_b=5$.\cite{ref8} 
The particle masses and lifetimes are taken from Ref. \citen{ref10}. 
The relevant CKM parameter values used are $V_{cb}=0.0395$,
$V_{cs}=1.04 $, $V_{cd}=0.224$, $V_{ud}=0.974$, $V_{us}
=0.2196$ and $V_{ub}=0.08\times V_{cb}=0.00316$. 
The decay constants are taken as $f_{\pi}=130.7$,
$f_K=159.8$; $f_{K^*}=214$;\cite{refa} $f_{\rho}=210$;\cite{refb} 
$f_D=220$, $f_{D^*}=230$, $f_{D_s}=240$, $f_{D_s^*}=260$\cite{refc} 
and $f_{a_1}=205$\cite{refd} (in MeV).
The decay constant $f_{J/\psi}$ is determined from the value of
$\Gamma(J/\psi \to e^+ e^- )$:\cite{ref10}
\begin{equation}
f_{J/\psi}=\sqrt{\frac{9}{4} \left (\frac{3}{4\pi \alpha^2}
\right ) \Gamma(J/\psi \to e^+ e^-)M_{J/\psi}}=404.5 ~\mbox{MeV}\;.
\end{equation}

The parameters $a_1$ and $a_2$ appearing in these decays, which have 
recently been determined from the CLEO data,\cite{ref0} are
$a_1=1.02$ and $a_2=0.23$.\cite{ref11}
Using these values we obtain
the branching ratios for $\bar B^0 $, $B^-$ and $\bar B_s$ mesons, which
are tabulated in Tables I, II and III, respectively. The overall
agreement between the predicted and experimental data is quite
remarkable. 
Here, it may be worthwhile to note that the relevant processes are
relativistic and the form factor functions $I(\omega )$ play 
significant roles: For example, the velocity of the final $D^+$
in the process $\bar B^0\to D^+\pi^-(\bar B^0\to D^+D_s^{*-})$
is $v_D=0.78c(0.70c)$, and the value of $I$ is $I=0.53(0.65)$.
The polarization and the angular distribution
parameters for $B \to D^* V$ decay modes are presented
in Table IV. The polarization fraction $(\Gamma_L/\Gamma$)
for $\bar B^0 \to D^{*+} \rho^- $ (0.883) agrees well with experimental
value $0.93 \pm 0.05 \pm 0.05 $.\cite{ref10} For the decay
mode $\bar B^0 \to \bar K^{*0}J /\psi $, $\Gamma_L/\Gamma $ (0.428)
also agrees with the
recent CLEO data $ 0.52 \pm 0.07 \pm 0.04 $.\cite{ref01}

In this paper we have calculated the branching ratios of
the exclusive nonleptonic two-meson decay of $B$ mesons using the
covariant oscillator quark model based on the factorization
approximation. The applied form factors are consistent with the
predictions of heavy quark symmetry.  The overall agreement of
our predictions for two meson nonleptonic decays of $B$ mesons with
the existing experimental data suggests that the factorization
approximation works well and the estimation of confinement effects
by the COQM is valid.

\acknowledgements

R. M. would like to thank CSIR, the Govt. of India, for a fellowship.
A. K. G. and M. P. K. acknowledge the financial support from
DST, the Govt. of India. The authors are grateful to Dr. K. Yamada
for informing them of the recent values of $\beta$ obtained from the 
analyses of mass spectra.

\end{document}